%% file: eprint.tex
\documentclass[12pt]{article}
\usepackage{graphicx}
\usepackage{hyperref}
\usepackage[activate={true,nocompatibility},final,tracking=true,kerning=true,spacing=nonfrench,factor=1100,stretch=15,shrink=15]{microtype}
\SetTracking{encoding={*}, shape=sc}{0}

\newcommand\pubnumber{}
\newcommand\pubdate{\today}

\def\institute{Centre for Cosmology, Particle Physics and Phenomenology\\
Universit\'e catholique de Louvain, B-1348 Louvain-la-Neuve, BELGIUM}
\def\support{\footnote{Supported by Fonds de la Recherche Scientifique (FNRS), Belgium.}}

\def\Title#1{\begin{center} {\Large #1 } \end{center}}
\def\Author#1{\begin{center}{ \sc #1} \end{center}}
\def\Address#1{\begin{center}{ \it #1} \end{center}}

\newcommand\pubblock{\rightline{\begin{tabular}{l} \pubnumber\\
         \pubdate  \end{tabular}}}
\newenvironment{Abstract}{\begin{quotation}  }{\end{quotation}}
\newenvironment{Presented}{\begin{quotation} \begin{center} 
             PRESENTED AT\end{center}\bigskip 
      \begin{center}\begin{large}}{\end{large}\end{center} \end{quotation}}

\input econfmacros.tex

\begin{document}
\begin{titlepage}
\pubblock

\vfill
\Title{Measurements of single top quark cross sections at 13~TeV with the CMS experiment}
\vfill
\Author{Matthias Komm\support, on behalf of the CMS Collaboration}
\Address{\institute}
\vfill
\begin{Abstract}
An overview of recent measurements of inclusive and differential single top quark cross sections at 13~TeV with the CMS experiment is given in this note. This includes measurements targeting the $t$-channel and tW production modes resulting in inclusive cross sections of $\sigma_{t\mathrm{\mbox{-}ch.}}=238\pm32~\mathrm{pb}$ and $\sigma_\mathrm{tW}=63.1\pm6.6~\mathrm{pb}$ respectively. In addition, the $t$-channel cross section has been measured differentially as a function of the top quark transverse momentum and rapidity. The results are found in agreement with the standard model expectations. Furthermore, a search for single top quark production in association with a Z~boson is detailed which yields an observed (expected) significance of 3.7 (3.1) standard deviations. 
\end{Abstract}
\vfill
\begin{Presented}
$10^{th}$ International Workshop on Top Quark Physics\\
Braga, Portugal,  September 17--22, 2017
\end{Presented}
\vfill
\end{titlepage}
\def\thefootnote{\fnsymbol{footnote}}
\setcounter{footnote}{0}

\section{Introduction}

The production mechanisms of single top quarks offer a unique experimental access to study details of electroweak interactions. For example, inclusive cross section measurements can be used to derive limits on the magnitude of the Cabibbo-Kobayashi-Maskawa (CKM) matrix element $\mathrm{V}_\mathrm{tb}$. Differential cross sections on the other hand allow to perform in-deep tests of the theoretical modelling and coupling structure. In this note recent measurements of single top quark production in $t$- and tW-channel as well as a search for tZq production at a centre-of-mass energy of 13~TeV with the CMS experiment~\cite{cms} are presented.

\section{$t$-channel}

\begin{figure}[!htb]
\begin{center}
\includegraphics[width=0.38\textwidth]{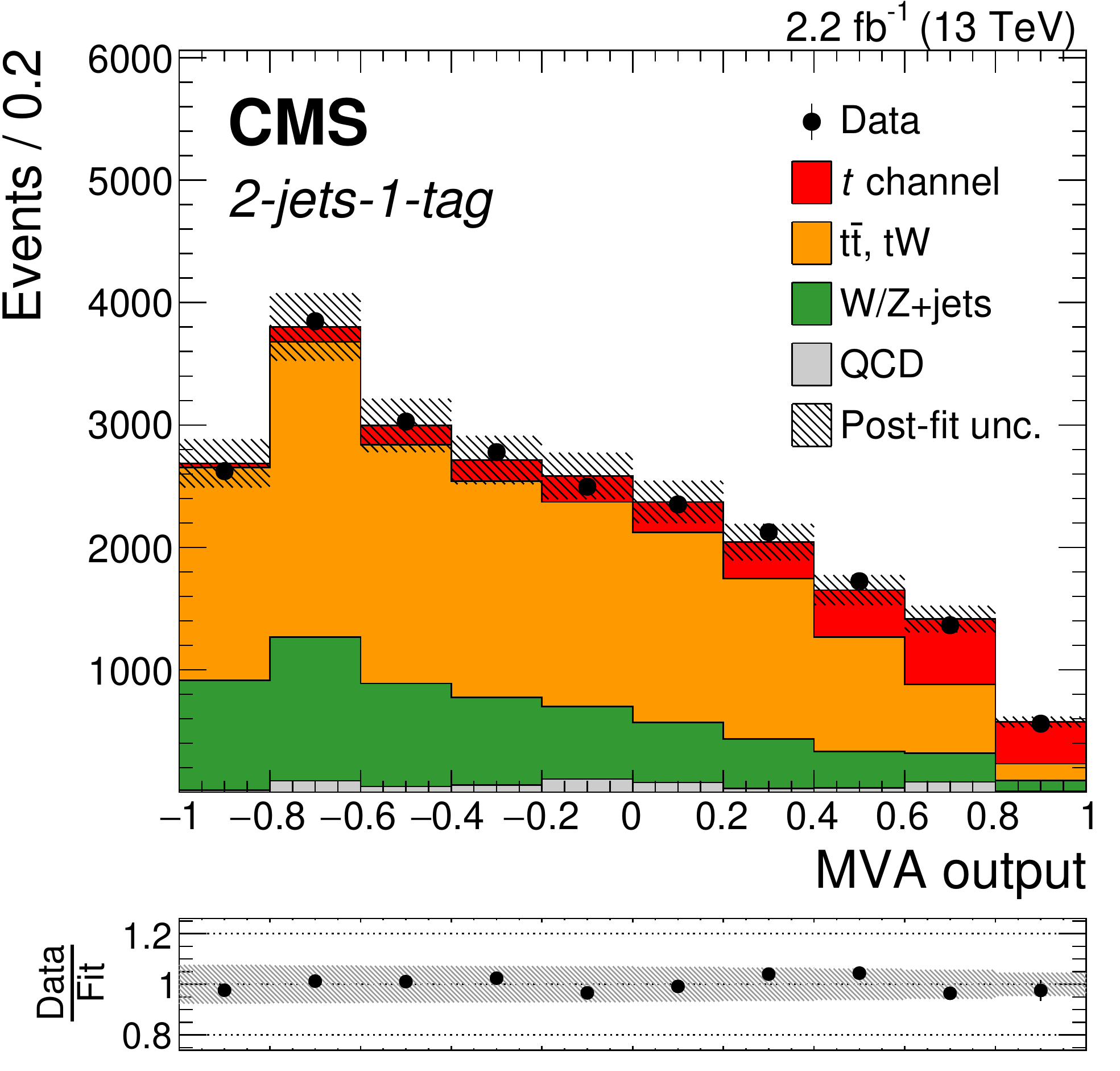}\hspace{0.02\textwidth}
\includegraphics[width=0.55\textwidth]{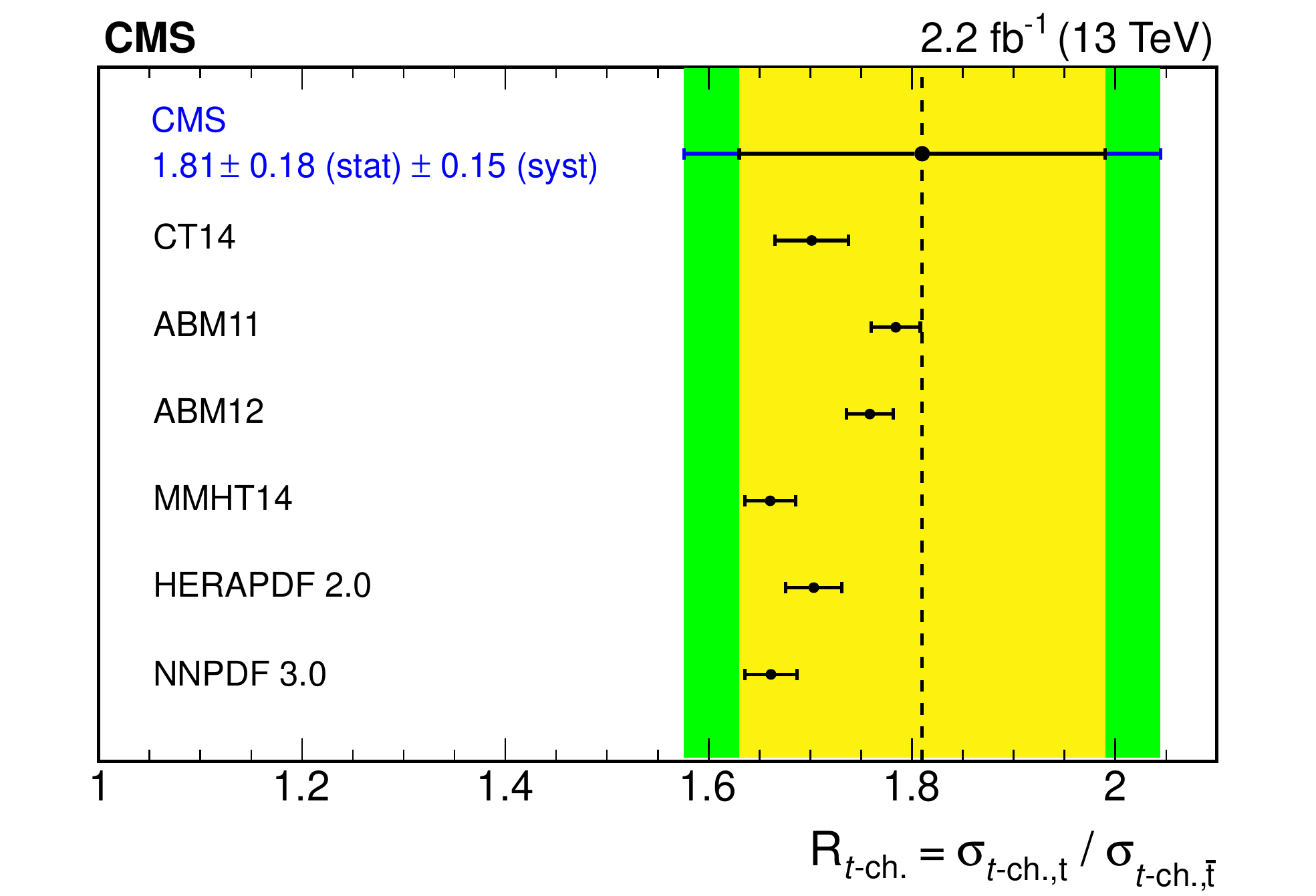}
\caption{\label{fig:t-channel}Measurement of inclusive $t$-channel cross section: (left)~distribution of neural network discriminant used for signal extraction; (right)~measured charge ratio compared to various PDF sets. The figures are taken from Ref~\cite{tchannel-inc}.}
\end{center}
\end{figure}

The inclusive and differential cross section of single top quark $t$-channel production is measured in proton-proton collision events containing one isolated muon, two or three jets, and significant missing transverse energy. The analysed dataset corresponds to $2.2~\mathrm{fb}^{-1}$. A multivariate discriminant is trained to separate signal from background events further. Its distribution in the signal region, consisting of two jets of which one is b-tagged (shown in Fig.~\ref{fig:t-channel}, left), and in $\mathrm{t}\bar{\mathrm{t}}$ control regions are used to estimate the amount of signal events through a maximum likelihood fit. An inclusive cross section of $\sigma_{t\mathrm{\mbox{-}ch.}}=238\pm32~\mathrm{pb}$~\cite{tchannel-inc} is measured which is well in agreement with the standard model expectation of $\sigma_{t\mathrm{\mbox{-}ch.}}^\mathrm{NLO}=217^{+9}_{-8}~\mathrm{pb}$~\cite{hathor}. Additionally the charge ratio, $R=\sigma(\mathrm{t})/\sigma(\bar{\mathrm{t}})$, is extracted by fitting the distributions for top quark and antiquark events separately. A comparison of the result with the predictions by various PDF sets is presented in Fig.~\ref{fig:t-channel} on the right.

\begin{figure}[!htb]
\begin{center}
\includegraphics[width=0.45\textwidth]{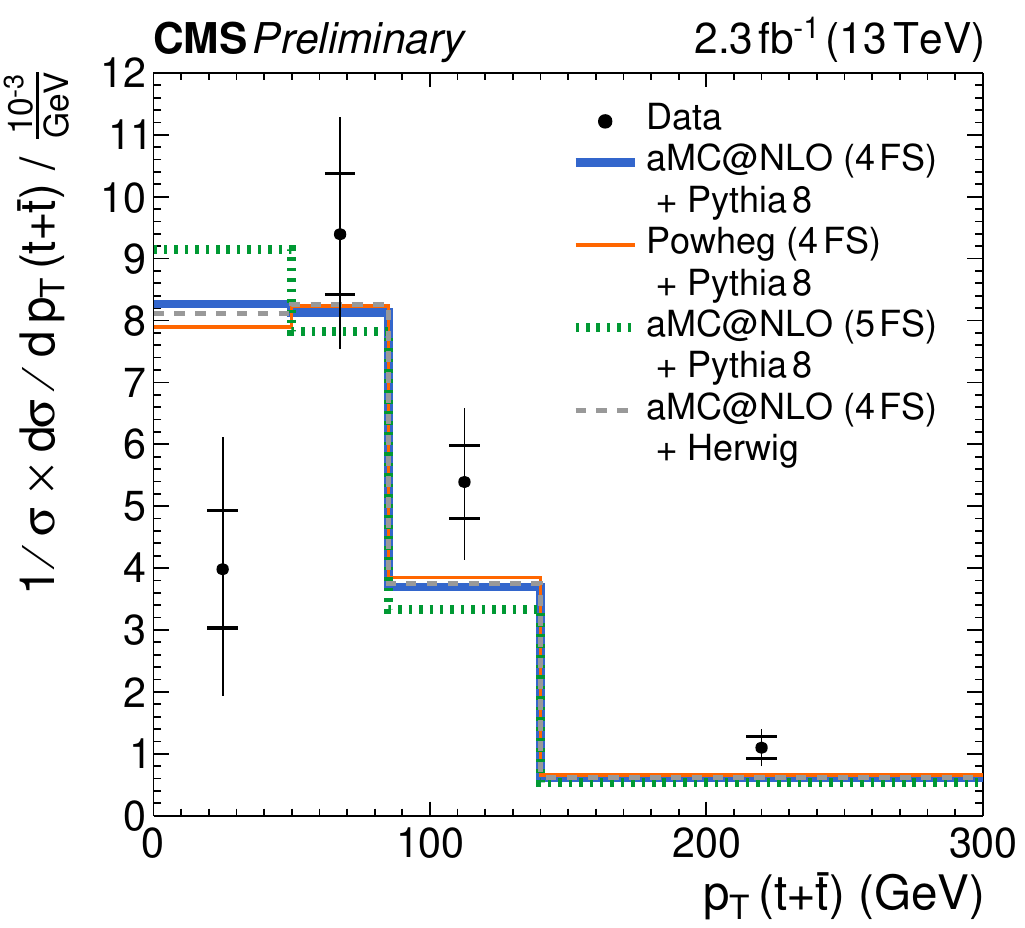}\hspace{0.02\textwidth}
\includegraphics[width=0.45\textwidth]{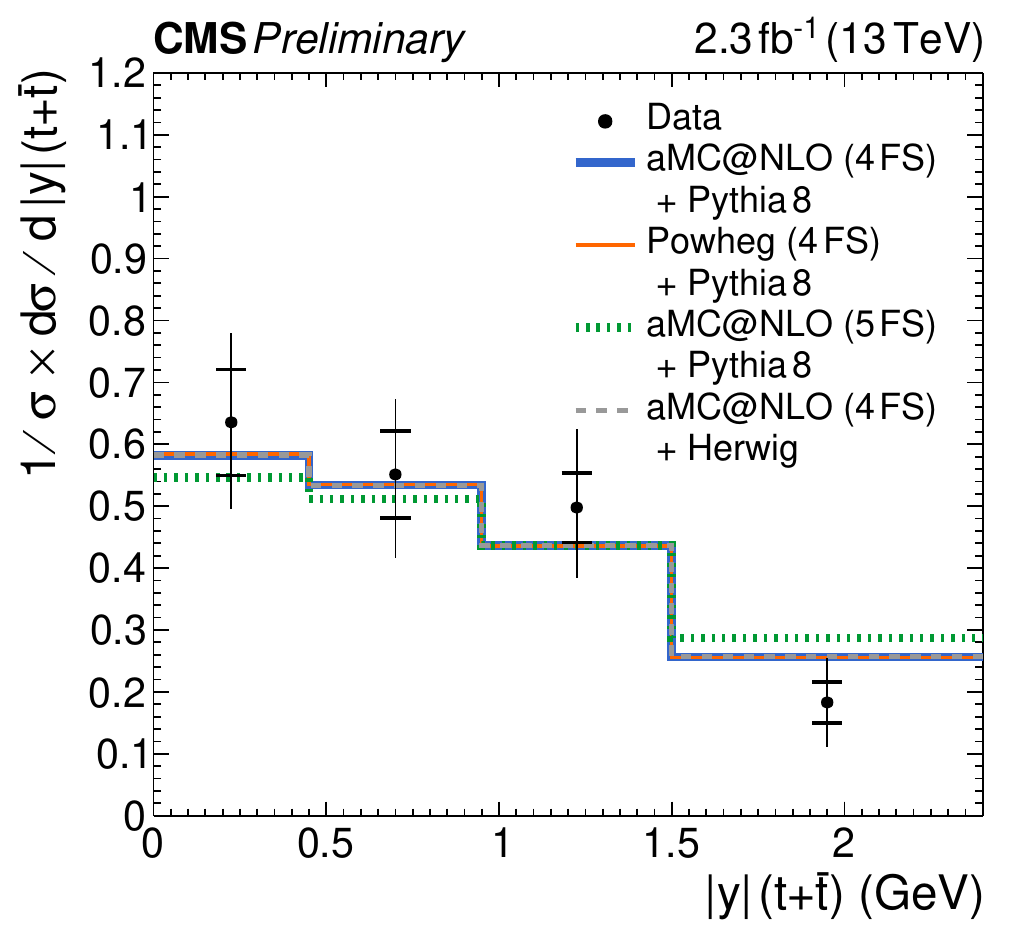}
\caption{\label{fig:tchan-diff}Differential $t$-channel cross section as a function of the top quark (left)~transverse momentum and (right)~rapidity. The figures are taken from Ref~\cite{tchannel-diff}.}
\end{center}
\end{figure}

The $t$-channel cross section is also measured as a function of the top quark transverse momentum and rapidity using a similar analysis strategy~\cite{tchannel-diff}. The results are presented in Fig.~\ref{fig:tchan-diff} and compared to the predictions by various event generators. Overall the results agree with the predictions within uncertainties.

\section{tW-channel}

\begin{figure}[!htb]
\begin{center}
\includegraphics[width=0.47\textwidth]{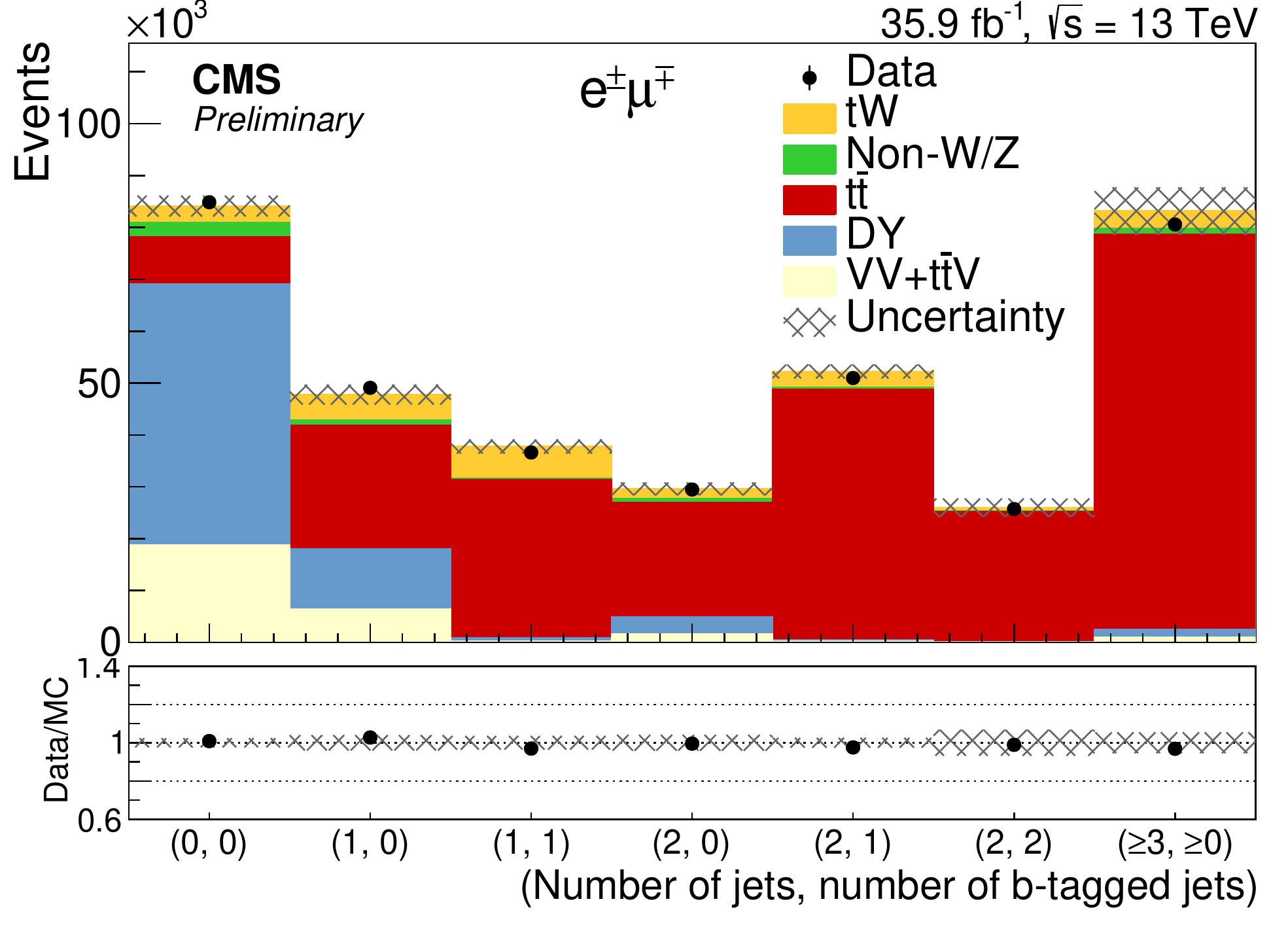}\hspace{0.02\textwidth}
\includegraphics[width=0.47\textwidth]{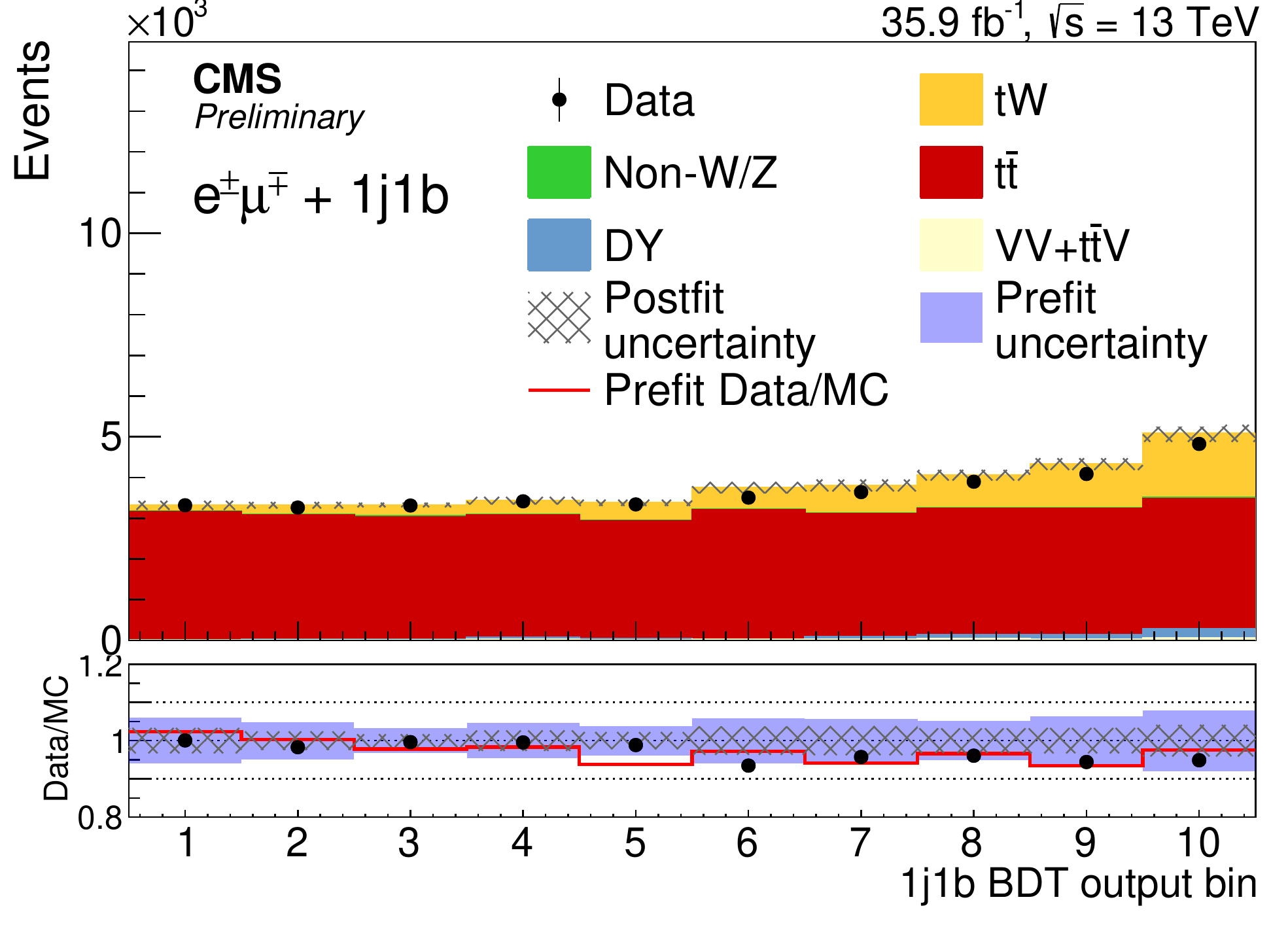}
\caption{\label{fig:tw}Single top quark cross section in tW-channel: (left)~predicted and observed number of event per category; (right)~distribution of the BDT discriminant in (1~jet, 1~b-tag)~category. The figures are taken from Ref~\cite{tw-inc}.}
\end{center}
\end{figure}

The inclusive cross section of single top quark production in association with a W~boson is measured in events containing two isolated leptons~($\mathrm{e}^{\pm}\mu^{\mp}$), jets, and significant missing transverse energy. The measurement is based on proton-proton collision data corresponding to $36~\mathrm{fb}^{-1}$. Events are categorised depending on the number of jets and the subset of jets which is also b-tagged as presented in Fig.~\ref{fig:tw} on the left. The cross section is estimated through a simultaneous maximum likelihood fit to the distributions of a trained boosted decision tree in the 1j1b~(shown in Fig.~\ref{fig:tw}, right) and 2j1b categories, and the distribution of the transverse momentum of the subleading jet in 2j2b $\mathrm{t}\bar{\mathrm{t}}$ control region. This results in a cross section of $\sigma_\mathrm{tW}=63.1\pm6.6~\mathrm{pb}$~\cite{tw-inc} which is found in agreement with the standard model expectation of $\sigma_\mathrm{tW}^\mathrm{NLO+NNLL}=71.1\pm3.9~\mathrm{pb}$~\cite{kidonakis}.

\section{tZq production}

A search for the production of single top quarks in association with a Z~boson is conducted using proton-proton data corresponding to $36~\mathrm{fb}^{-1}$. Events containing three leptons (electrons/muons), two or three jets, and significant missing transverse energy are selected. A high separation of signal from background events is achieved by combining various observables into a powerful discriminant using a boosted decision tree. In particular, various probabilities that a certain event stems either from the signal process or from the $\mathrm{t}\bar{\mathrm{t}}\mathrm{Z}$ or WZ+jets background processes are used as inputs. These are calculated using the matrix element method in which the matrix elements of a given process are integrated over the final state momenta while accounting for the resolution of reconstructed objects with respect to their partonic counterparts through so-called transfer functions.

\begin{figure}[!htb]
\begin{center}
\includegraphics[width=0.97\textwidth]{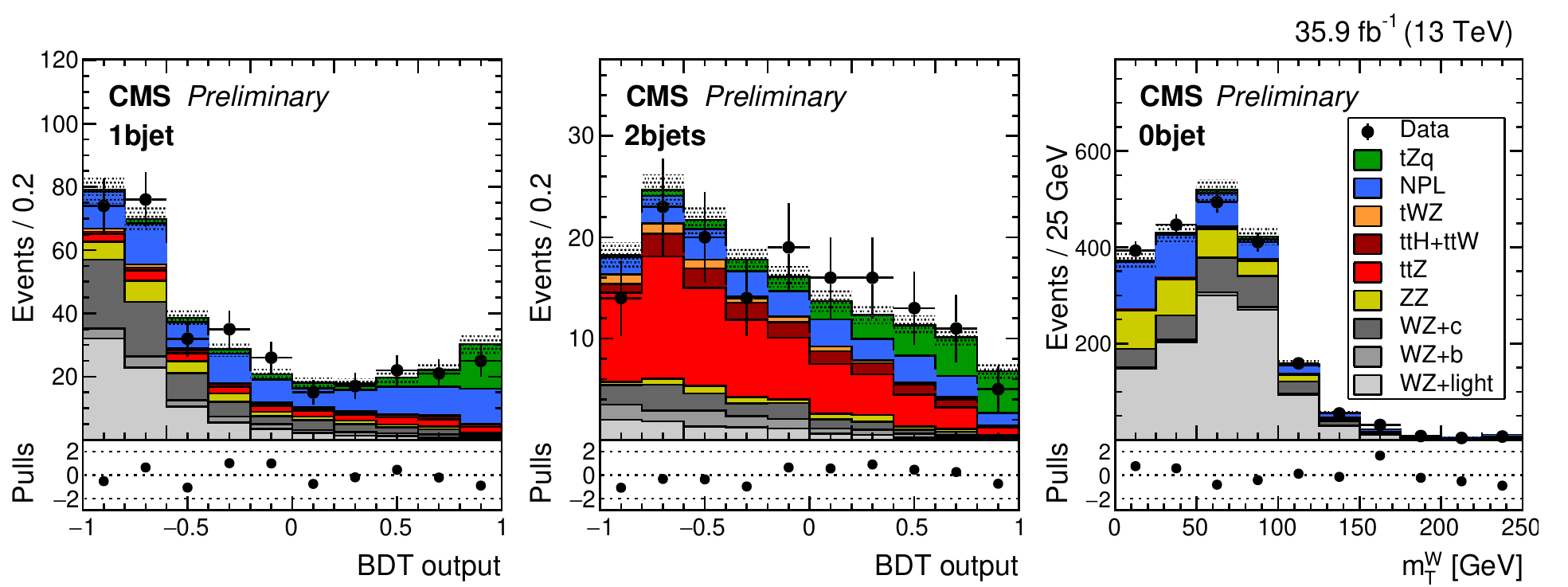}
\caption{\label{fig:tzq}Distributions used in the maximum likelihood fit for estimating the tZq cross section: boosted decision tree discriminant in (left)~1~b-tag and (middle)~2~b-tag region; (right)~transverse W~boson mass in 0~b-tag region. The figures are taken from Ref~\cite{tZq-inc}.}
\end{center}
\end{figure}

A cross section of $\sigma_\mathrm{tZq}=123^{+44}_{-39}~\mathrm{fb}$~\cite{tZq-inc}, where the Z~boson decays leptonically, is measured through a maximum likelihood fit using events with no, one, or two b-tagged jets as displayed in Fig.~\ref{fig:tzq}. Compared with the standard model expectation of $\sigma^\mathrm{NLO}_\mathrm{tZq}=94\pm3~\mathrm{fb}$ (calculated with MC{@}NLO~\cite{mcatnlo}) this represents an observed significance of 3.7 standard deviations while a significance of 3.1 standard deviations is expected.

\section{Conclusion}

Inclusive single top quark cross sections in the $t$- and tW-channel have been measured using proton-proton collision data at 13~TeV with the CMS experiment. Additionally, the $t$-channel cross section has also been measured differentially. Furthermore a search for tZq production has been carried out resulting in an evidence with an observed significance of 3.7 standard deviations. Overall, the presented results are found in agreement with the standard model expectations.

\end{document}

%% file: econfmacros.tex
\def\beq{\begin{equation}}
\def\eeq#1{\label{#1}\end{equation}}
\def\eeqn{\end{equation}}

\def\beqa{\begin{eqnarray}}
\def\eeqa#1{\label{#1}\end{eqnarray}}
\def\eeqan{\end{eqnarray}}

\let\bar=\overbar

\def\Dslash{\not{\hbox{\kern-4pt $D$}}}
\def\dslash{\not{\hbox{\kern-2pt $\del$}}}

\def\msb{{\bar{\ssstyle M \kern -1pt S}}}

